# Cloud and IoT based Smart Agent-driven Simulation of Human Gait for Detecting Muscles Disorder

Sina Saadati[1], Mohammadreza Razzazi[2]

**Abstract**

Motion disorders pose a significant global health concern and are often managed with pharmacological treatments that may lead to undesirable long-term effects. Current therapeutic strategies lack differentiation between healthy and unhealthy muscles in a patient, necessitating a targeted approach to distinguish between musculature. There is still no motion analyzer application for this purpose. Additionally, there is a deep gap in motion analysis software as some studies prioritize simulation, neglecting software needs, while others concentrate on computational aspects, disregarding simulation nuances. We introduce a comprehensive five-phase methodology to analyze the neuromuscular system of the lower body during gait. The first phase employs an innovative IoT-based method for motion signal capture. The second and third phases involve an agent-driven biomechanical model of the lower body skeleton and a model of human voluntary muscle. Thus, using an agent-driven approach, motion-captured signals can be converted to neural stimuli. The simulation results are then analyzed by our proposed ensemble neural network framework in the fourth step in order to detect abnormal motion in each joint. Finally, the results are shown by a user-friendly graphical interface which promotes the usability of the method. Utilizing the developed application, we simulate the neuromusculoskeletal system of some patients during the gait cycle, enabling the classification of healthy and pathological muscle activity through joint-based analysis. This study leverages cloud computing to create an infrastructure-independent application which is globally accessible. The proposed application enables experts to differentiate between healthy and unhealthy muscles in a patient by simulating his gait.

## 1. Introduction

Movement is an essential factor in everybody's life and affects diverse aspects of people. However, a significant proportion of people are suffering from motion disorders affecting over a billion patients worldwide [1]. Thus, controlling of motion disorders is an important field in science, technology, and clinical studies [2-4]. Motion analysis plays a vital role in detecting motion disorders and in their subsequent treatment and control [4-6].

Motion is a complex phenomenon in the human body which originates from the contribution of three compound systems including neural, muscular, and skeletal known as the neuromusculoskeletal system [7]. This process begins from the neural system in which motion commands are sent to the muscles through motor neurons [7-10]. Muscles generate a contraction force as a reaction to neural stimulations [5-10]. The muscular force is then applied to the skeletal system which behaves as a mechanical lever and, finally, movement happens [5-6]. Any failure in each segment of the neuromusculoskeletal system causes a motion disorder [7]. For instance, common disorders such as multiple sclerosis (MS) or Parkinson disease are arising from of neural failures in which muscles cannot be stimulated correctly [10-12].

[1] Department of Computer Engineering, Amirkabir University of Technology (Tehran Polytechnic), Tehran, Iran. sina.saadati@aut.ac.ir
[2] Department of Computer Engineering, Amirkabir University of Technology (Tehran Polytechnic), Tehran, Iran. razzazi@aut.ac.ir

Medications are widely used to treat or control the disorder symptoms including chemical drugs [13-15]. Considering the disadvantages of chemical medications with their long-term harmful effects, a critical issue in treatment procedures is that while only a small proportion of muscles are affected by disorders, all the muscles are uniformly influenced by medications, even muscles of the eyes for example [14-17]. Consequently, it is required to differentiate between healthy and unhealthy muscles in motion disorder treatment.

Since many common movement disorders, such as Parkinson disease or MS, are based on the neural failures, it is necessary to modeling the motion in such a way in which neural activity patterns in muscles can be calculated by an inverse dynamic method in order to distinguish between healthy and unhealthy muscles in these patients. One important challenge of modeling process is maintaining the naturalness of the phenomenon. In the context of human motion and the gait, naturalness of modeling means that the entities desinged in the modeling process must accurately mimic the physiological, neurological, and biomechanical details as they occur inside of the human body [88-90]. This approach ensures that the modeling process is interpretable, the results obtained are reliable, and the causes of any phenomena can be precisely tracked and explained [45,79].

From a physiological perspective, in many studies such as [67-73] in which human motion is modeled, muscles are modeled as atomic indivisible modules. However, in the human body, muscles are composed of elements called motor units (MUs), that can be independently stimulated and contracted. Notably, according to Henneman's size principle, motor units are stimulated with different patterns to produce the required force with minimal energy consumption [4-6,10,87]. Thus, if motor units are not considered as the constituent elements of muscles, a significant portion of the details related to muscle's physiological behavior will be omitted from the modeling process, making it impossible to analyze movement at the level of individual MUs. Considering MUs in muscle modeling not only preserves the naturalness of the simulation but also allows for the differentiation and analysis of the various physiological patterns of MUs.

In addition to distinguishing muscles by their constituent motor units, it is required that the physiological function within these MUs be accurately modeled to maintain naturalness. This requires detailed modeling of all stages of a MU's contraction, from receiving the neural signal to producing mechanical force which includes the pattern of increasing and decreasing contraction force, the effect of increased neural stimulation frequency on contraction force, and the state of tetanus (when the stimulation frequency is intense and the MU produces maximum possible force). Many muscular phenomena, such as the manner of increasing or decreasing contraction intensity, or muscle weakening in cold environments or because of continuous activity, occur due to physiological changes within MUs [5-6,10]. Therefore, if the behavioral and physiological details of MUs are neglected in modeling, it will be impossible to scrutinize and justify many muscular behaviors in human movement.

Attention to the principle of naturalness not only ensures the modeling process is simple and comprehensible but also makes it more intuitive. In some studies such as [72-73,84-86], different models have been proposed for different muscles. Besides the challenge of muscle model diversity, these models are often limited to certain environmental conditions such as the individual's gender, age, or even ambient temperature. The diversity and limitations of muscle models make movement modeling and analysis in individuals very complex and time-consuming. In studies using machine learning for muscle modeling, two additional challenges arise: lack of interpretability and data dependency. Despite functional differences in muscles, all voluntary muscles in humans, from powerful muscles like the hamstrings in the thigh to delicate muscles such as those in the eyes and eyebrows, share the same physiological structure [4-6,10]. Thus, if the principle of naturalness is adhered to in muscle modeling, a comprehensive and flexible model can be provided to explain all muscle behaviors across different muscle groups and environmental conditions.

From a neurological perspective, neural messages in the human body are transmitted through electrochemical reactions known as action potentials (APs), which travel along neurons. APs, regardless of the type of message they carry, all have the same structure. Therefore, they are considered as discrete pulses inside the body [10-12]. This concept is sometimes equated by mistake with electromyographic (EMG) signals. EMG is an invasive method in which electrodes are inserted into muscles, attempt to estimate the intensity of neural pulse transmissions by recording electrochemical changes in a muscle [5-6,74]. Consequently, the nature of EMG signals is different from that of APs. Thus, to maintain the naturalness f the modeling, the concept of APs should be considered as neural messages instead of EMG signals. This allows for a more accurate modeling of the relationship between the neural and muscular systems, enhancing the interpretability of neuromuscular system modeling.

From a biomechanical perspective, biomechanical models are used to connect motion signals with neuromuscular behavior modeling as the human skeletal system functions like a complex lever [5]. According to the laws of biomechanics, the skeletal system can alter the amount of force applied to the bones; for example, in some cases like the elbow or knee joint, it can increase movement speed by reducing muscle contraction force, and in other cases like the heel, it can amplify the force produced by the muscle [5-6]. To simulate the biomechanics of the human body, the skeletal system is modeled as a geometric graph where bones are modeled as edges and joints as nodes [5-6]. During walking, each of the lower limb joints independently converts muscular forces into kinematic factors such as force, acceleration, and speed. Therefore, to maintain the principle of naturalness, each joint must be modeled as an agent. In this way, joint agents can connect to muscle agents and model the movement as it occurs inside of the human body.

To maintain the principle of naturalness in biological and medical modeling, it is suggested to follow an agent-based approach [18,21-24]. The agent-based modeling method considers body structures as composed of independent and autonomous elements called agents that can interact with each other in parallel [19-23]. In the context of human movement, all muscles and MUs, neurons and neural message transmission, joints, and the transmission of biomechanical forces in bones occur simultaneously [10]. Neglecting this approach in motion modeling results in overlooking the parallel nature of living systems' activities in the human body. However, if body parts are modeled as independent agents, the principle of naturalness is maintained, and it also becomes possible to use parallel computations for human movement simulation. If parallel execution of modeling elements is feasible, computational speed is enhanced, and in more complex modeling, that deeper levels of details are considered, distributed computations can be utilized.

Although naturalness in modeling is necessary, collecting the required data for simulation input can still be a challenge. In many methods, wearable sensors are used as tools for gathering movement signals [60-66]. However, access to these sensors is a limiting factor as many people around the world cannot afford these tools. This challenge encompasses various economic and social dimensions. In many regions with more resources, purchasing or producing wearable sensors can incur economic costs. In other regions, due to technological and social limitations, access to wearable sensors is still impossible, even with significant expenditure. Therefore, techniques are required to improve the availability and accessibility of human motion measurement tools. To this end, an Internet of Things (IoT)-based method is proposed in this paper as smart phones are more prevalent than wearable sensors, even in many underprivileged areas.

Facilitating the collection of necessary data for simulation input by IoT devices does not simplify the entire simulation process. Simulation software, such as OpenSim, AnyLogic, or AnyBody, typically requires a complex process of installation, deployment, execution, and use [77,79]. Many of these simulators are not independent of the infrastructure, which limits them to running on a limited set of computer infrastructures. This makes the use of simulation tools challenging for users. Thus, a process must is required in which simulation tools are independent of hardware and software infrastructures to facilitate their execution and use by users. To this end, the concept of cloud computing is utilized in this paper, whereby simulations are run on a fully cloud-based platform.

By creating a simulation on a cloud-based platform and using IoT facilities, it is possible to run the simulator anywhere in the world with many types of computers. Although in many studies, the analysis and interpretation of simulation outputs are entrusted to physicians and relevant specialists [63-73], it is possible to use artificial intelligence and deep learning methods to analyze simulation results. This approach can assist specialists in analyzing simulation outputs and enhance the accuracy of their analyses. Since clinical specialists work in stressful environments that can reduce their diagnostic accuracy over time, the need for tools to compensate for this decline in accuracy is more pronounced. However, caution must be exercised when using neural network and deep learning methods in medical applications. Despite their tremendous computational power and learning capabilities, these methods can carry the risk of lacking interpretability and explainability. In medical applications, it is crucial that all results and outputs of deep learning models are as traceable, interpretable, and explainable as possible to justify the cause of each output and , consequently, the outputs will be more reliable.

Although AI-based models can analyze simulation outputs, the results they provide may not be understandable to ordinary users. To complement the analyzing process, it is necessary to present the system's results and outputs to users through a user-friendly interface. A powerful graphical interface allows all the computational complexities of the proposed method, including the simulation of various movement systems in the human body and deep learning analysis to be covered from the user, displaying only the final results in a way that is easily understandable. Consequently, a user-friendly interface should minimize the interactions between the user and the software and be comprehensible to ordinary users. From a clinical perspective, specialists working in stressful environments need minimal required interactions with the software. From the perspective of non-specialist users, simplicity and comprehensibility of the software will attract more people to use it, thereby accelerating the attention given to health care by the general public.

In similar studies, there are significant gaps in modeling discussions, software requirements, and more. From the perspective of the importance of natural interpretability, none of the previous studies have modeled the whole muscles and joints as independent agents, which can undermine the naturalness of the simulations. In many of these studies, EMG signals have been incorrectly equated with neural messages, which also violates the naturalness and interpretability of the simulations. The differentiation of MU functions within a muscle during a contraction has also been overlooked by many studies. From the perspective of software requirements such as accessibility and usability, similar studies have used markers or wearable sensors, limiting their use to specific laboratory conditions. The use of cloud capabilities in the development of simulators has also been neglected. Additionally, many simulators are not independent of infrastructure for execution, which limits their installation and use. Attention to user-friendly interface criteria, such as minimizing the interaction between human and computer in order to ease its learning and use, has also been overlooked in many studies.

This paper presents a method for analyzing human movement and distinguishing between healthy and unhealthy muscles. In this method, movement signals are first collected using smartphone sensors based on an IoT-based approach. Then, using the collected data, human gait is simulated with an agent-based method to reconstruct and analyze the biomechanical and neurophysiological aspects of movement in a software environment. In the movement modeling process, muscles are modeled based on their constituent elements, i.e., MUs. Additionally, neural messages are reconstructed by modeling APs. Muscles and joints are modeled as independent agents, which allows movement to be simulated in the software environment as it occurs in the human body. The simulation results are then presented to specialists through appropriate graphical user interfaces and are also analyzed by an ensemble deep learning-based framework to assist specialists in interpreting the simulation results. Efforts have been made to maximize interpretability in the deep learning framework. Finally, using cloud computing, software was developed in which all computations, including simulations and neural network models, are executed on the cloud. This approach makes the tool infrastructure-independent, allowing it to be accessible and usable worldwide, even in many underprivileged and poor areas with limited technological resources. In this tool, simulation results and neural network model outputs are presented to users through a beautiful user interface that is easy for the general public to learn. The designed user interface minimizes interactions between the user and the system, making it understandable not only for specialists but also for ordinary users.

Thus, the contribution of this study is as follows:

- **IoT method for motion signals capturing:** A methodology is proposed in which motion signals can be captured using sensors of mobile devices easily. This approach significantly enhances the usability of the proposed method for movement modeling compared to methods requiring markers or wearable sensors.
- **Agent-driven biomechanical model of the human lower body:** An agent-based model is proposed in which the muscular forces can be calculated from angular displacements of lower body joints during a gait. Lower body joints are modeled as agents to maintain the naturalness of the simulation.
- **Agent-base model of human voluntary muscles:** An agent-base model of human voluntary muscle is proposed which can be used inside of the proposed biomechanical model. Neural stimulations of each muscle can be calculated by this model. In this paper, neural messages are modeled as APs, and muscles are modeled as a collection of independent MUs. This approach, while maintaining the naturalness of the modeling, enables specialists to analyze movement at the deep level of individual MUs.
- **Deep learning framework for signal processing:** An ensemble neural network framework is proposed in order to analyze the motion of lower body joints and detect if the gait is normal or pathological. In this paper, biomechanical, muscular, and neural patterns during the gait are independently examined by three separate neural network models. This approach, as opposed to using a single complex neural network with ambiguous performance, will bring greater reliability and interpretability to the results.
- **Design and development of user-friendly cloud-based application:** An open-source application is developed that can capture motion signals through a mobile device and analyze it. The application is infrastructure-independent as can be executed on a wide range of computers.

In Section 2, preliminaries are explained which are the basis of our proposed method. Related works are also surveyed briefly in this section. In Section 3, our methodology for analysis of human motion during gait is detailed. In Section 4, our proposed method is analyzed in order to investigate its precision, reliability, and interpretability. In Section 5, the discussion and our future works are described.

# 2. Literature Review

In this section, the preliminaries of our proposed method are described. In Section 2.1, the agent-driven method for modeling and simulation is described. Then, the physiology and biology of the human voluntary muscles as the main motors of the motion are described in Section 2.2. Section 2.3, describes how artificial intelligence models and software techniques can promote analyzer applications. Finally, some related works are investigated in Section 2.4.

## 2.1. Agent-based Modeling and Simulation (ABMS)

In many modeling and simulation processes, while the appearance of simulation in the computer is the same as what is occurring in the real world, there is a significant gap between the algorithm that is executed inside the simulator and the procedure which is happening inside nature [18-20]. ABMS methodology suggests that the simulations be natural meaning that the algorithm inside the simulation should be equal or, at least, similar to the procedure of phenomenon inside nature [19-20].

Based on the agent-driven method for modeling, the world is formed of autonomous elements, named agents, which have an internal decision-making system despite conventional simulation methods in which all the elements are controlled by an exterior component. In an agent-driven simulation, different types of agents can be defined and also some agents can form the structure of other agents. In addition, agents can interact with each other, and with their environment [19-21]. For instance, Figure 1 illustrates the difference between agent-driven and traditional methods for modeling a humanized society in which lines signify data streams.

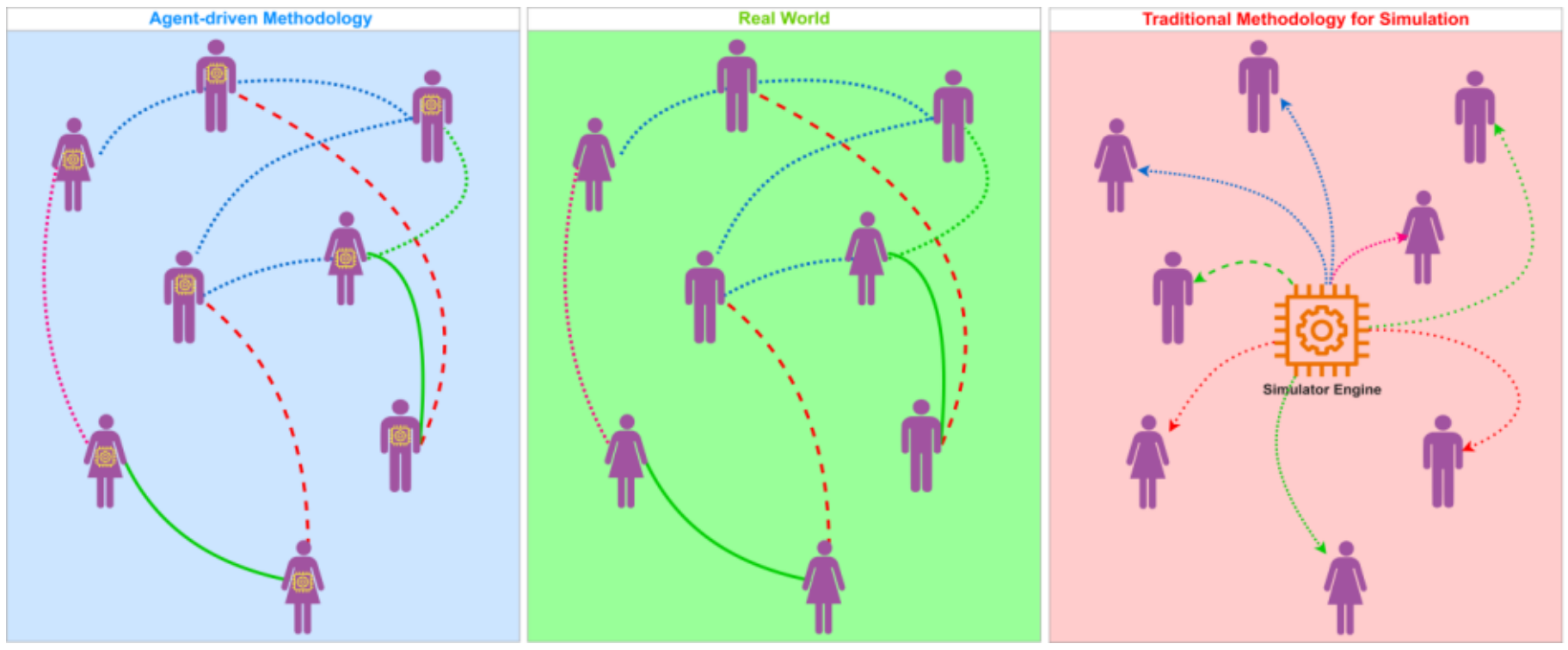

Figure 1. Comparison between agent-driven and traditional modeling of human society.

By following the principles of ABMS, each aspect of a phenomenon should be reflected in the simulation. Thus, the modeling will be more interpretable [18-19]. As a consequence, the results of the simulation will be more reliable and robust. Since interpretability is a critical factor in clinical applications and scientific research, this method is recommended to be used in medical and biological simulations [18,22-24].

## 2.2. Physiology and Biology of Muscles and Motor Control

The motion process begins from the neural system in which movement commands are sent to the muscles. The motion commands are electrochemical interactions, known as Action Potentials (APs), that are moved forward along the motor neurons [10,25-26]. After the muscle is stimulated by APs, chemical interactions happen inside the muscles which turn the chemical energy of Adenosine Triphosphate (ATP) into mechanical energy [5-6,10,25]. Finally, a physical force is generated and applied to the skeletal system [5-6].

Muscles are known as the motors of the motion in both humans and animals. A particular group of muscles, known as voluntary muscles, can be consciously controlled and are attached to the bones [5-6]. These muscles are composed of elongated bundles called fascicles, bound together by perimysium. Each fascicle contains a number of muscle fibers that are ensheathed together by endomysium. Perimysium and endomysium are connective tissue sheaths that encompass fascicles and muscle fibers, respectively, providing structural support and facilitating the transmission of force during muscle contraction. The muscle fibers are biologically cells that contain nuclei, mitochondria, and a number of myofibrils covered by sarcoplasmic reticulums [4-6,10,27].

Each myofibril contains numerous macromolecules named actin and myosin filaments. Once an AP is received, calcium ions are released from the sarcoplasmic reticula into the myofibrils. This release initiates chemical interactions between actins and myosins, resulting in the contraction of the myofibrils, which collectively results in the contraction of the entire muscle. In this structure, a number of muscle fibers are connected to a single motor neuron [4-6,10]. This network is known as the Motor Unit (MU) and follows a one-to-many relationship, meaning that all the muscle fibers within a unit are stimulated and controlled by only one motor neuron [5-6].

In this study, we present a novel approach by modeling and simulating APs instead of EMG signals. This method enhances the naturalness and interpretability of the simulation. Additionally, for the first time, muscles are modeled based on their MUs with precise connections to the neural system, rather than using an ambiguous black-box module. The size principle, which dictates how MUs are activated within muscles, is incorporated into our inverse dynamic simulation, allowing for MUs stimulations differentiation. Furthermore, our muscle model is grounded in the physiological structure of muscles, enabling comprehensive modeling and investigation of all muscles, from those in the lower body to the eye muscles, using our proposed agent-driven model.

While muscles are motors of human motions, the skeleton plays a critical mechanical role in the kinematics of the movements. For instance, the contraction force of the gastrocnemius muscle will be first intensified in the ankle joint and then applied to the heel. Consequently, the biomechanical impact of the skeletal system must be considered in the motion analysis. Biomechanics is a research field in which the mechanical aspects of the human and animal bodies are studied [5].

Different movements exhibit varying levels of complexity and hold distinct degrees of importance in daily life. Consequently, within the realm of human motion analysis, certain exercises are deemed more significant due to their functional mobility. For instance, walking which is referred to as the gait cycle is an important movement in biomechanical and clinical analysis. It contains the actions which take place in a single leg during walking including the stance phase, when the feet are located on the ground, and the swing phase, when the leg is being moved forward to take a step [28-29]. Gait is also important in motion analysis and disorder detection [6,29]. Thus, in this study, the human gait is concentrated to be simulated and analyzed using an agent-driven methodology.

## 2.3. Artificial Intelligence and Softwares Requirements

Machine learning is a subfield of artificial intelligence that computers extract knowledge from data [30-32]. Artificial Neural Networks (ANNs) are a subfield of machine learning models that are implemented based on the learning process inside the human brain. ANNs are computationally powerful as they are able to realize complex patterns in compound data [33-37]. Thus, they are widely used in medical and clinical applications to assist physicians in order to simplify the medical processes, enhance accuracy, and reduce costs [38-41].

ANNs are layered networks of perceptrons which map the input data to the output results. In this architecture, the perceptron is a mathematical model of a single human neuron and plays the main role in the learning of the models. Each perceptron receives multiple numbers as its input, then calculates a weighted sum of all inputs, and then, using an activation function, produces an output value. The output value can then be sent to other perceptrons [33-34,37].

By increasing the number of perceptrons and layers of an ANN, it can realize more complex patterns from data. Thus, this field of computing is also referred to as deep learning [34-35]. However, the computations of ANNs with a huge number of parameters are hard to be explained or interpreted so that these models might not be reliable in medical processes [42-45,82-83]. Nevertheless, the challenge can be remedied by providing interpretable methods in which ensemble ANN models are developed instead of one compound and ambiguous model [42] .

In an ensemble model, multiple ANNs are used, each focusing on one aspect of the phenomenon. Following this method, not only interpretability is enhanced, but also the risk of noise detection is minimized as all the ANN models learn equal patterns while learning different noises. As a consequence, there is no overlap between the noises, and a reliable and interpretable result will be calculated. This phenomenon is known as cross-training and is considered in some studies to increase the accuracy of the models [46-47].

While ANNs are very powerful and advantageous in learning compound patterns, they are complex concepts in computer science and cannot be used by inexpert people. Thus, software engineering techniques are used to cover the complexity of computations and make intelligent applications transparent and useable for users. As a result, it is important that modern applications have a user-friendly graphical interface so that even inexpert users can learn the application and use it easily [48-52].

In addition to the user-friendly interface, modern applications should be infrastructure-independent so that they can be easily executed on a wide range of computers, with no difference between a PC or mobile device, and without any compound installation procedure. Because of this, cloud-based applications are getting popular nowadays as they are not only infrastructure-independent, but also available, and economically effective [53-57]. IoT is another software technology which can promote the power of applications. Mobile devices are significantly considered in IoT methods owing to their precise sensors and computing powers [53]. Hence, applications can be more useful, available, and cost-effective if they connect to mobile devices and use their functionalities [53-59].

We have utilized IoT and cloud computing to enable users to employ the proposed methodology from any location. Unlike wearable sensors, which may not be available even in some metropolises such as Isfahan, our proposed IoT method for motion signal capturing is accessible worldwide, including in rural areas with minimal technological infrastructure. The fully cloud-based platform allows users to execute our application without the need for installation, regardless of their hardware or software configurations. Consequently, even residents of rural areas can access our advanced analysis application using a smartphone.

## 2.4. Related Works

The simulation of human muscles and motion analysis of motion represents a captivating area of study and research. Within this section, we conduct a brief comparative analysis of related studies, examining their limitations from simulation, artificial intelligence, and software engineering points of view. Our comparison is detailed in Table 1.

OpenSim and AnyBody are powerful tools built for the investigation of the biomechanical behavior of human motion. They enable experts to analyze different muscles within different motions [60-62]. However, the software engineering requirements such as user-friendly interface or infrastructure-independency are not considered in those works making them challenging for users.

A cloud-based method is proposed by [63] to analyze and monitor the patient's motions in which neural networks are utilized. Nevertheless, as a subset of signal processing operations are executed in the local computer, the method is not fully cloud-based and, as a consequence, not infrastructure-independent.

Machine learning models are proposed by [64-66] in which motion signals are captured from insole-based methods or mounted accelerometers and then, abnormal movements are detected. Nevertheless, the models are not interpretable as physiological and biological aspects of motion are not considered. Additionally, signal capturing could be facilitated by an IoT-based approach.

An application is proposed by [67] in order to detect motor impairment severity in patients suffering from Parkinson's disease. Data collection is done by the screen of mobile. However, the method is restricted to be used only on iPhone mobiles. The machine learning models are not interpretable.

Other artificial intelligence methods for the analysis of the gait are proposed by [68-69] that are limited in the classification of gait phases. There is no proposed abnormality detection in these methods. In addition, data are received through insoles or electrogoniometers which are not accessible devices in many regions.

From a flexibility point of view, the models proposed by [70-71] can be used for a wide range of muscles and animals. However, interpretability is not considered as the biology of the neuromuscular system is not modeled.

EMG signals are analyzed in proposed methods by [72-73]. Although, the algorithms are not natural as APs are the real neural stimulations inside the human body which are far from what can be seen in electromyograms [8,74]. Therefore, they have used signal processing techniques for EMG signals which breaches the principle of naturalness of the modeling. The methods are also restricted to analyzing the motions of only a single joint. The Naturalness of simulation also enables modelers to simulate environmental conditions such as temperature or genetics. This is because the effect of these factors is justified by their influence on the biological features of the muscles [5-6,10].

Table 1. Comparison of the human motion analysis methods, assuming full satisfactory by ✓, partial satisfactory by ☐, and failing to satisfactory by ×.

| Metrics | | [60] | [62] | [63] | [64] | [65] | [66] | [67] | [68] | [69] | [70] | [71] | [72] | [73] |
|---|---|---|---|---|---|---|---|---|---|---|---|---|---|---|
| Differentiation between healthy and unhealthy muscles (RQ1) | | × | × | × | × | × | × | × | × | × | × | × | × | × |
| Natural Modeling and Simulation | Agent-driven method for modeling | × | × | × | × | × | × | × | × | × | × | × | × | × |
| | AP-based simulation of neuromuscular system | × | × | × | × | × | × | × | × | × | × | × | × | × |
| | Physiological simulation of the muscle | ✓ | × | × | × | × | × | × | × | ☐ | × | ☐ | × | ☐ |
| | Ability to simulate muscular environmental conditions | ✓ | ✓ | × | × | × | × | × | × | × | × | × | × | × |
| | Flexibility of the model | □ | □ | × | × | × | × | × | × | × | ✓ | ✓ | × | × |
| Software Requirements (RQ3) | User-friendly graphical interface | × | × | × | × | × | × | ✓ | × | × | × | × | × | × |
| | Cloud-based platform | × | × | □ | × | × | × | × | × | × | × | × | × | × |
| | Hardware and software infrastructure independency | × | × | × | × | × | × | × | × | × | × | × | × | × |
| | IoT-based approach | × | × | × | × | × | × | □ | × | × | × | × | × | × |
| Artificial Intelligence (RQ4) | Pathological gait detection | □ | × | ✓ | ✓ | ✓ | ✓ | × | × | × | × | × | × | × |
| | Interpretability of the method | × | × | × | × | × | × | □ | × | × | × | × | □ | × |
| | Cross-training learning | × | × | × | × | × | × | × | × | × | × | × | × | × |

# 3. Methodology

In this section, our methodology for gait analysis is detailed. An overview of the proposed method is demonstrated in Figure 2. Motion signals are first captured by sensors of mobile devices (Step 1). Then, muscular forces are calculated (Step 2). Then, the APs entered into each MU can be computed by the muscle agents (Step 3). Finally, simulation results are analyzed by deep-learning models (Step 4) and displayed to the users by a user-friendly interface (Step 5).

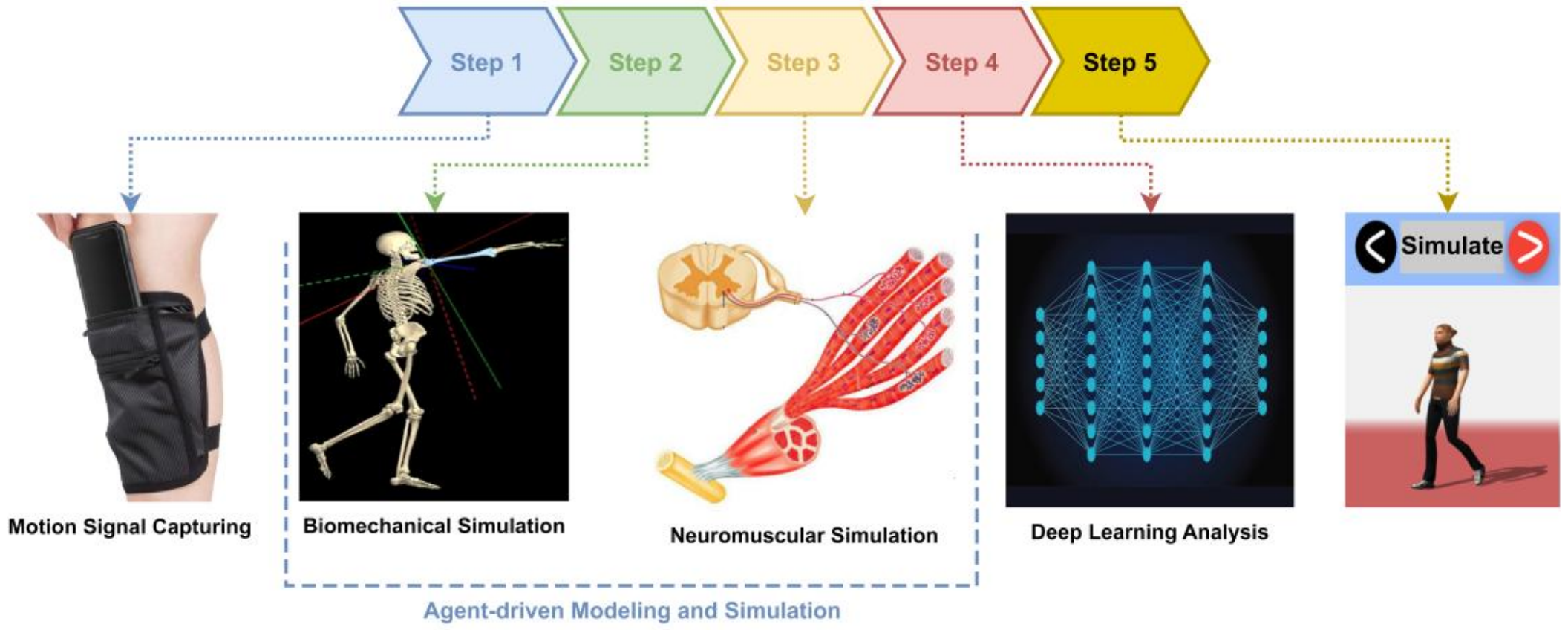

Figure 2. Overview of the proposed methodology for gait analysis.

Section 3.1 details the motion signals capturing method. The agent-driven biomechanical model of the human lower body is described in Section 3.2. The agent-based model of human voluntary muscle is explained in Section 3.3 as well as deep learning methods for analysis of results that are detailed in Section 3.4.

## 3.1. Proposed IoT-based Method to Capture the Motion Signals

Due to the complexity, expensiveness, and not availability of wearable sensors, an IoT method is proposed in this study to capture human motion. Mobile devices contain a 3D compass that can detect the orientation of the device on the ground by providing three values for the angles of the device relative to the three axes of rotation. The compass is sensitive to small motions of the device and is widely used in applications such as games, traveling, astronomy, artificial reality, and augmented reality.

In this study, a web-based application is developed which can connect to the 3D compass of the mobile and capture the motions during gait. The application provides a user-friendly graphical interface which can be even used by inexpert users. When users click on the capturing icon, ten seconds are given to users to put the mobile in the organ's position. After that, a warning sound is played and the user has 12 seconds to walk. Finally, another warning sound is played to inform the user to stop walking.

Following the capturing phase, the raw data during the walking is processed. The raw data is a matrix of angles during the movement. First, noisy values in the matrix are detected and removed. Then, a mathematical regression model is used to summarize all the steps into a single gait cycle matrix. Additionally, some features such as minimum and maximum angle values, number of steps, stance/swing time proportion, speed, and time per step are calculated which are valuable in subsequent analysis. The process is illustrated in Figure 3. The output of the regression model is used for the agent-driven simulation while feature extraction output provides additional and valuable information for further deep learning analysis.

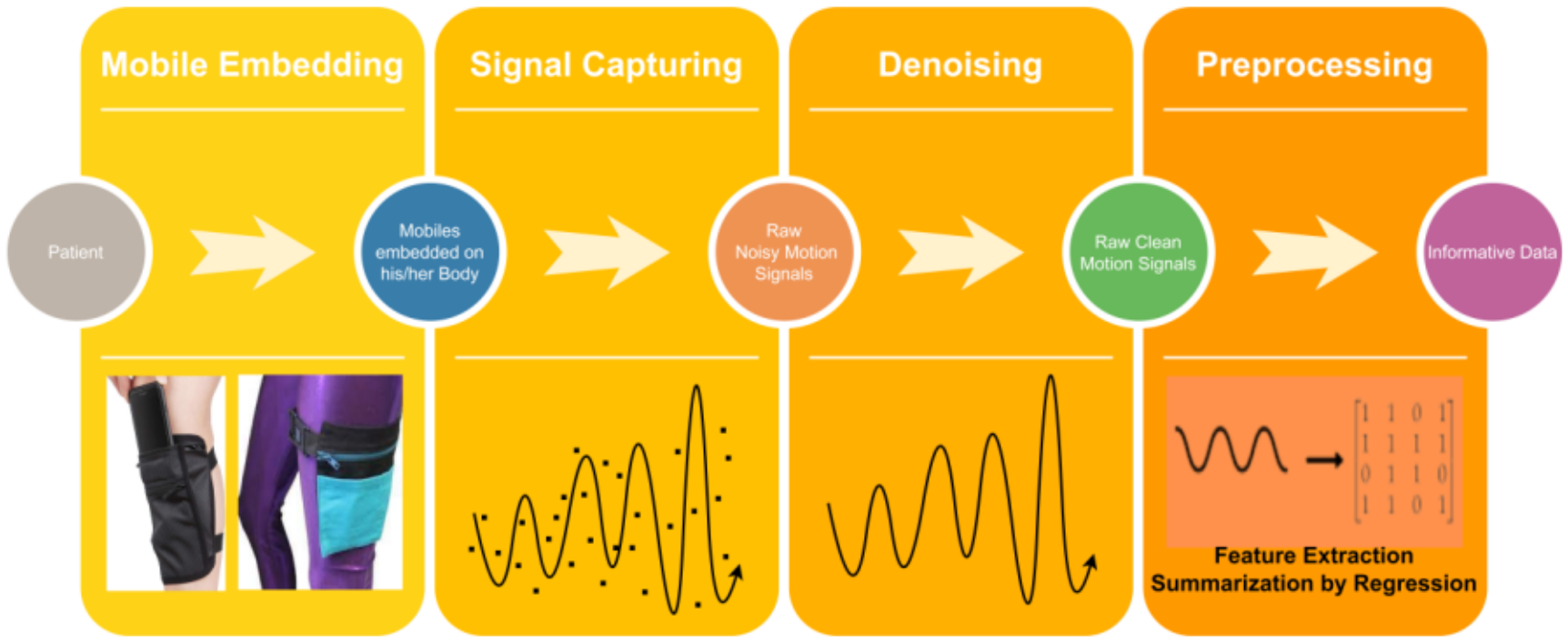

Figure 3. Proposed IoT-based method for motion signal capturing.

### 3.2. Proposed Agent-driven Biomechanical Model of Lower Body

To accurately analyze human motion throughout a gait cycle, it is essential to employ a biomechanical model in order to convert angular displacements of the joints into muscular contraction forces. Thus, this section proposes an agent-driven biomechanical model for the lower body, governed by the proposed Boots algorithm. Figure 4 presents our innovative agent-driven model of the human lower body. Muscle agents that are assembled in this model are described in Section 3.3.

Our agent-driven model for the human lower body encompasses three joint agents and six muscle group agents: the Gastrocnemius, Tibialis Anterior, Hamstrings, Quadriceps, Gluteus Maximus, and Iliopsoas. Given that the primary function of the Gastrocnemius muscle is plantar flexion, we consider its influence on the knee joint to be negligible for this study, thereby simplifying the computational framework. Additionally, the gait is investigated as a 2D motion happening in the sagittal plane to simplify the analysis.

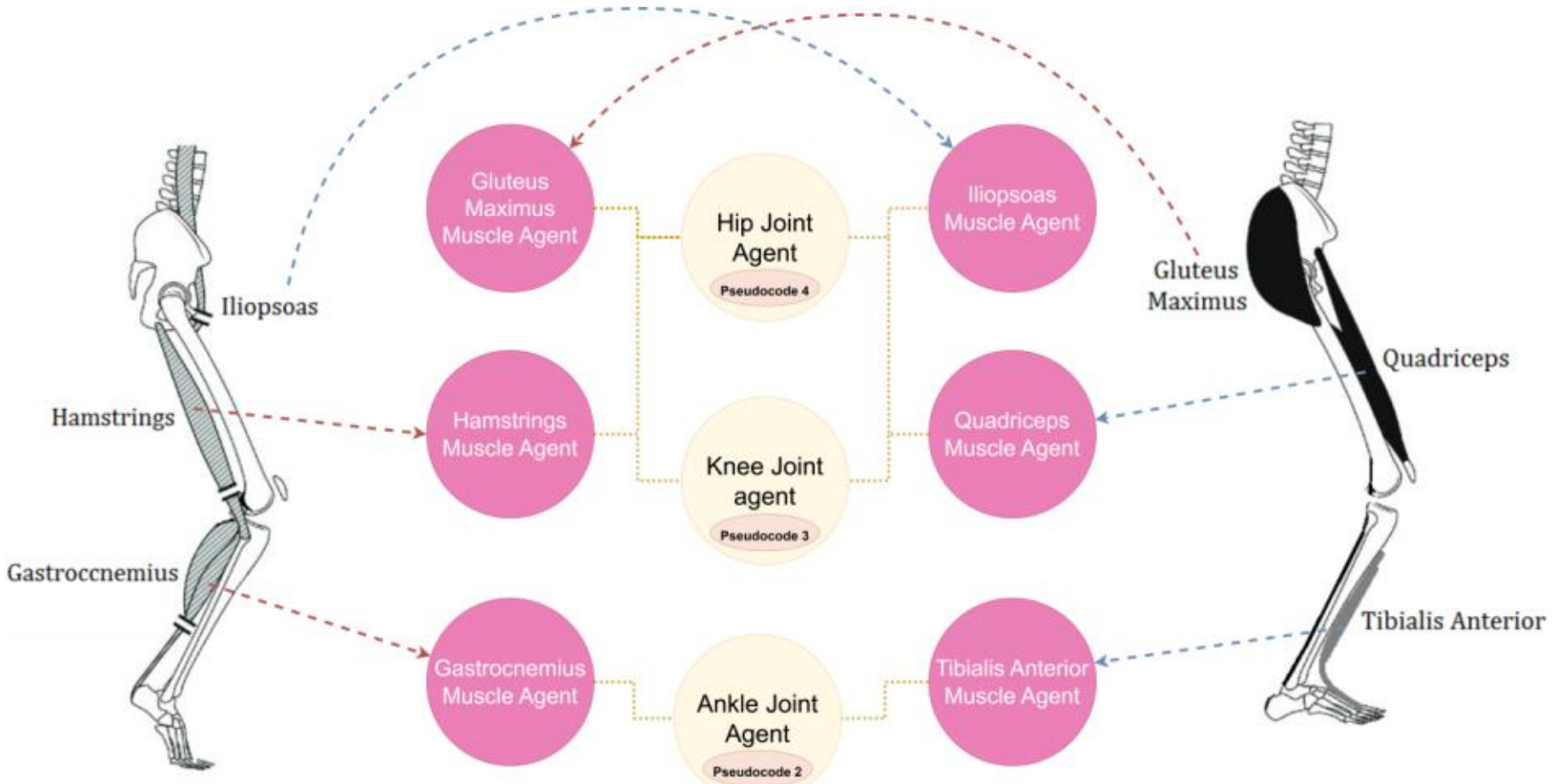

Figure 4. Our proposed agent-driven biomechanical model for lower body.

To accurately characterize the biomechanical behavior of our agent-based model, we introduce the Boots algorithm. The annotations for the algorithm are provided in Table 2 in which $O$, $J$, $t$, $H$, $V$, and $P$ represent organ and joint of interest, the current time, horizontal or vertical direction, and an anatomic point, respectively. Annotations listed in Table 2 can be defined as variables or matrix during the development of the application. For example, each annotation that varies over time can be defined as an array of values representing a matrix.

Table 2, Definition of annotations in the Boots algorithm.

| Feature | Equation Statement | Derivative | Details |
|---|---|---|---|
| Mass | $M^{(O)}$ | $M^{(body)}$ | Mass of the patient |
| | | $M^{(feet)}$ | $0.0145 \times M^{(body)}$ [5] |
| | | $M^{(leg)}$ | $0.0465 \times M^{(body)}$ [5] |
| | | $M^{(thigh)}$ | $0.100 \times M^{(body)}$ [5] |
| Weight | $W_{(t)}^{(J)}$ | $W_{(t)}^{(ankle)}$ | The felt weight in the ankle joint. |
| | | $W_{(t)}^{(knee)}$ | The felt weight in the knee joint. |
| | | $W_{(t)}^{(hip)}$ | The felt weight in the hip joint. |
| Length | $L^{(O)}$ | $L^{(feet)}$ | Distance from ankle joint to toe. |
| | | $L^{(leg)}$ | Distance from ankle to knee. |
| | | $L^{(thigh)}$ | Distance from knee to hip joint. |
| Moment of Inertia | $I^{(O)}$ | $I^{(ankle)}$ | $M^{(feet)} \times 0.62^2 \times {L^{(feet)}}^2$ [5] |
| | | $I^{(knee)}$ | $M^{(leg)} \times 0.528^2 \times {L^{(leg)}}^2$ [5] |
| | | $I^{(thigh)}$ | $M^{(thigh)} \times 0.540^2 \times {L^{(thigh)}}^2$ [5] |
| Angular Acceleration | $\alpha_{(t)}^{(J)}$ | $\alpha_{(t)}^{(ankle)}$ | $\alpha_{(t)} = \frac{d}{dt} v_{(t)} = \frac{d^2}{dt^2} \theta(t)$ [5] |
| | | | $\alpha_{(t)} = \frac{\Delta}{\Delta t} v_{(t)} = \frac{\Delta^2}{\Delta t^2} \theta(t)$ [5] |
| Ground Reaction Force | $F_{ground}^{(P)}(t)$ | $F_{ground}^{(toe)}(t)$ | Computable using $d_{(bodyCentre\ ,\ toe)}^{(horizon)}$ |
| | | $F_{ground}^{(heel)}(t)$ | $W_{(t)}^{(ankle)} - F_{ground}^{(toe)}(t)$ |
| Angle | $\theta_{(H)}^{(O)}(t)$ | $\theta_{(H)}^{(feet)}(t)$ | The horizontal angles for the lines connecting the ankle to the toe, the ankle to the knee, and the knee to the hip joints, respectively. |
| | | $\theta_{(H)}^{(leg)}(t)$ | |
| | | $\theta_{(H)}^{(thigh)}(t)$ | |
| | | $\theta_{(V)}^{(feet)}(t)$ | The vertical angles for the lines connecting |
| | $\theta^{(O)}_{(V)}(t)$ | $\theta^{(leg)}_{(V)}(t)$ | the ankle to the toe, the ankle to the knee, and the knee to the hip joints, respectively. |
| | | $\theta^{(thigh)}_{(V)}(t)$ | |
| Distance | $d^{(H/V)}_{(P_a\,,P_b)}\,(t)$ | $d^{(H)}_{(bodyCentre\,,\ toe)}(t)$ | The horizontal distance between mass centre of the body and toe of the feet. |
| | | $d^{(H)}_{(bodyCentre\,,\ hip)}(t)$ | The horizontal distance between mass centre of the body and hip joint. |
| | $d_{(P_a\,,P_b)}$ | $d_{(ankle\,,\ toe)}$ | The distance from ankle to toe. |
| | | $d_{(heel\,,\ ankle)}$ | The distance from heel to ankle. |
| | | $d_{(ankle\,,\ feetCentre)}$ | The direct distance from mass centre of the feet to the ankle. |
| | | $d_{(ankle\,,\ knee)}$ | The distance from knee to ankle. |
| | | $d_{(knee\,,\ legCentre)}$ | The distance from the centre of the leg and knee. |
| Torque | $\tau^{(J)}_{(t)}$ | $\tau^{(ankle)}_{(t)}$ | $I^{(ankle)} \times \alpha^{(ankle)}_{(t)}$ |
| | | $\tau^{(knee)}_{(t)}$ | $I^{(knee)} \times \alpha^{(knee)}_{(t)}$ |
| | | $\tau^{(hip)}_{(t)}$ | $I^{(hip)} \times \alpha^{(hip)}_{(t)}$ |

The Boots algorithm calculates the contraction force generated by each muscle group during the gait cycle. This calculation is achieved by determining the total torque resulting from the joint's angular acceleration. Then, the algorithm differentiates the torque generated by the muscle of interest from other factors, which are defined as environmental torques in this study.

Boots algorithm is the key functionality of joint agents , which takes three inputs: the agent of the joint in question, the agent of the muscle group whose contraction force is being calculated, and the environmental torques present at each step of the gait cycle, represented as a 2D array. Environmental torques encompass all torques caused by forces other than those from the target muscle group, which include the weight of body parts, ground reaction forces, and forces from non-target muscles.

Boots algorithm initiates by creating a zero-filled list (line 2). A 'For' loop then processes each gait cycle time step (lines 3-12), calculating the joint's total torque from its angular acceleration at that time step (line 4). The environmental torques are then deducted to isolate the muscle's contraction torque (lines 5-7). The algorithm concludes by computing the contraction force from the torque, recording the outcome (lines 8-11).

Given that muscles can only contract, a negative torque value calculated in Line 10 suggests interference from other muscles. Therefore, the contraction force for the muscle of interest is set to zero using the *ReLU* function.

Boots Algorithm( Joint , MuscleOfInterest , EnvironmentalTorques [ ] [ ] )<br>
Input: Agent of Joint, Agent of muscle group of interest, list of all the environmental torques.<br>
Output: Diagram of contraction force generated by group muscle of interest.<br>
1 **Begin**<br>
2 Results = Array with zero values.<br>
3 **For** *time* **in** Gait Cycle Diagram<br>
**Begin**<br>
4 TotalTorque = $\tau^{(J)}_{(t)}$<br>
5 **For** $i$ **from** 0 **to** EnvironmentalTorques[ $t$ ]. length( )<br>
6 TotalTorque −= EnvironmentalTorques [ $t$ ] [ $i$ ]<br>
7 $\theta$ = the angle between MuscleOfInterest contraction force and the bone.

8 $Dist$ = The distance between Joint and the ligament of the MuscleOfInterest .
9 ContForce $= ReLU\left(\frac{\text{TotalTorque}}{Dist \times Sin(\theta)}\right)$
10 Results [ time ] = ContForce
11 **End**
12 **Return** Results
13 **End**

Gastrocnemius Algorithm describes how the Boots algorithm can be applied to the ankle joint agent to calculate contraction forces of the gastrocnemius muscle. The primary function of the gastrocnemius muscle is to support body weight in the ankle and facilitate the body's forward acceleration. Thus, the torque from the ground reaction force—reflecting the body weight in the feet— (Lines 2-4), and the torque caused by the weight of the feet (Line 5) are considered as environmental torques. Finally, the contraction force is calculated using the Boots algorithm (Line 6). In this study, array concatenation is performed using the 'Concat' function.

Gastrocnemius Algorithm( Ankle Joint )
Input: Agent of the ankle joint.
Output: Diagram of physical force generated by the group muscle of gastrocnemius.
1 **Begin**
2 Torques1 = $F_{ground}^{(toe)}(t) \times d_{(ankle\ ,\ toe)}^{(H)} \times Sin\left(\theta_{(V)}^{(feet)}(t)\right)$
3 Torques2 = $F_{ground}^{(heel)}(t) \times d_{(heel\ ,\ ankle)}^{(H)} \times Sin\left(\theta_{(V)}^{(feet)}(t)\right)$
4 EnTorques1 = Torques1 − Torques2
5 EnTorques2 = $-M^{(feet)} \times 9.8 \times d_{(ankle\ ,\ feetCentre)} \times Sin\left(\theta_{V}^{(feet)}(t)\right)$
6 **Return** BootsAlgorithm( AnkleAgent , GastrocnemiusMuscleAgent , Concat( EnTorques1 , EnTorques2 ) )
7 **End**

The value of $F_{ground}^{(toe)}(time)$ can be determined by analyzing how the center mass of the entire body is maintained throughout the gait cycle. To facilitate this analysis, Pseudocode A1 has been developed and is included in Appendix A.

The contraction force generated by the quadriceps muscle agent can be calculated as shown in Quadriceps Algorithm. The quadriceps has two tasks: Negating the body weight in the knee joint and holding the position of the body (lines 2-3), and holding the body position at the back and hip joint (lines 4-5). In each case, environmental torques are defined based on biomechanical tasks of the muscle. By considering these tasks, the contraction force of the muscle can be calculated.

Quadriceps Algorithm( Knee Joint , Hip Joint)
Input: Agents of the knee and hip joint.
Output: Diagram of physical force generated by the group muscle of quadriceps(Quads).
1 **Begin**
2 EnTorques1 = $W_{(t)}^{(ankle)} \times d_{(knee\ ,\ ankle)} \times Sin\left(\theta_{(V)}^{(leg)}(t)\right)$
3 KneeForces = BootsAlgorithm( AnkleAgent , QuadsMuscleAgent , EnTorques1 )
4 EnTorques2 = $W_{(t)}^{(hip)} \times d_{(bodyCentre\ ,\ hip)}^{(H)}(t)$
5 HipForces = BootsAlgorithm( AnkleAgent , QuadsMuscleAgent , EnTorques2 )
6 **Return** (KneeForces + HipForces)
7 **End**

Gluteus Algorithm outlines the calculation of the contraction force of the gluteus maximus muscle in the hip joint agent. The principal function of the gluteus maximus is hip extension. Therefore, it is necessary to account for the torques resulting from the ground reaction force at the feet (Line 2) and the weight of the leg from the hip to the toes (Line 3). Furthermore, the torques generated by the hamstrings and quadriceps muscles are also considered (Lines 4-5), due to their impact on the hip joint. By considering these torques as environmental, the contraction force of the muscle can be calculated using the Boots algorithm (Line 6).

The muscular agents for the Tibialis Anterior, Hamstrings, and Iliopsoas have tasks similar to those of the Gastrocnemius, Quadriceps, and Gluteus Maximus, respectively. Therefore, their behaviors can be modeled using the proposed strategy. For instance, for the tibialis anterior muscle, the variables EnTorque1 and EnTorque2 should be considered with opposite signs as defined in Pseudocode 2. Additionally, the Hamstrings Algorithm mirrors the Quadriceps Algorithm, differing only in angular orientation.

Gluteus Algorithm(Knee Joint , Hip Joint)
Input: Agents of the hip and knee joint.
Output: Diagram of physical force generated by the group muscle of gluteus maximus.

1 **Begin**

2 EnTorques1 = $W_{(t)}^{(hip)} \times \left(L^{(leg)} + L^{(thigh)}\right) \times Sin\left(\theta_{(V)}^{(thigh)}(t)\right)$

3 EnTorques2 = $\left(M^{(thigh)} + M^{(leg)} + M^{(feet)}\right) \times d_{(hip\ ,\ fullLegCentre)} \times Sin\left(\theta_{(V)}^{(thigh)}(t)\right)$

4 EnTorques3 = $+$ Hamstrings Contraction Forces(Knee Joint , Hip Joint) $\times\ L^{(thigh)} \times Sin(5°)$

5 EnTorques4 = $-$ Gastrocnemius Contraction Forces(Ankle Joint ) $\times\ L^{(thigh)} \times Sin(5°)$

6 **Return** BootsAlgorithm( AnkleAgent , GastrocnemiusMuscleAgent , Concat( EnTorques1 , EnTorques2 , EnTorques3 , EnTorques4 ) )

7 **End**

Utilizing the Boots algorithm and the proposed agent-driven biomechanical model, the kinematics of motion can be used to ascertain the tension forces exerted by the lower body's muscle groups. Then, an agent-based model of muscles needs to be assembled in the biomechanical model in order to calculate the muscular neural stimulation (APs) from muscular contraction forces. Thus, the agents of muscles are detailed in Section 3.3.

## 3.3. Proposed Agent-driven Model of Human Voluntary Muscles

In this section, our proposed agent-based model for human voluntary muscle is delineated. The architecture of our model is meticulously crafted to mirror the biological and physiological intricacies of muscular structure. The model enables us to convert the contraction forces of MUs to the neural simuli (APs) for further analysis. Figure 5 depicts the proposed agent-based model for human voluntary muscle.

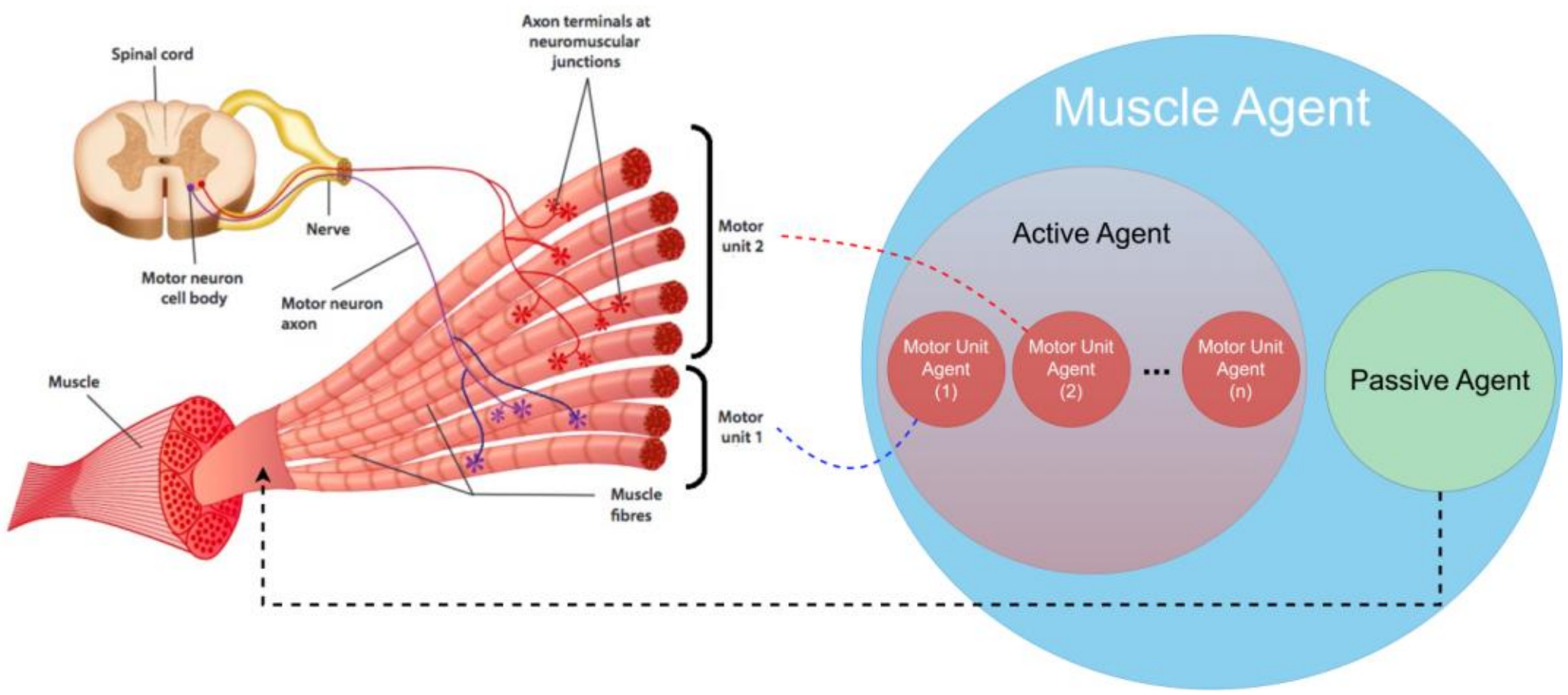

Figure 5. The proposed agent-based model of the muscle.

In this study, each muscle is investigated as an agent. Each muscle agent comprises two inner agents: active and passive agents. The active agent, composed of MU agents, acts in response to nervous stimuli, namely APs. The number of MU agents within the active agent is determined by the muscle category. For instance, muscles in the lower body have a greater number of MUs compared to facial muscles.

A formula is introduced by [5] to elucidate the force generated by a single MU receiving a solitary stimulus or AP. The formula is delineated in Equation 1, where $F_0$ represents a constant value that depends on the muscle category, $T$ denotes the time interval from the stimulation to the peak tension of the MU, which depends on the MU type and environmental conditions, and $t$ signifies the time.

$$F_{winter}(t) = F_0 \frac{t}{T} e^{\left(\frac{-t}{T}\right)} \qquad (1)$$

Equation 1 is derived from the chemical interactions within muscle fibers, which are universally equal across all types of voluntary muscles. Consequently, this equation enables the modeling of contractions for any MU under various environmental conditions. For instance, factors such as ambient temperature or muscular fatigue can affect the variable $T$ in Equation 1, thereby altering contraction dynamics.

MU contractions range from 10 milliseconds to 100 milliseconds. However, Equation 1 suggests that a standard MU in the lower body muscles may sustain contraction for over 600 milliseconds following a single stimulus. To address this discrepancy, we have formulated a new equation presented as Equation 2. This equation offers a refined analysis of MU behavior when receiving a singular AP. The parameters $F_0$, $T$, and $t$ retain their definitions from Equation 1.

$$F_{singleStimuli}(t) = F_0 \frac{t}{T} t^{\left(\frac{-t}{T}\right)} \qquad (2)$$

Equation 2 quantifies the tension force generated by a MU in response to a single AP. If subsequent stimuli are applied before the muscle has fully relaxed the cycle of calcium ion release within the myofibrils is initiated again. Due to the incomplete obliteration of calcium ions from the previous stimulus, an accumulation occurs within the myofibril, enhancing the interaction between actin and myosin filaments and thereby intensifying the contraction. This effect, termed wave summation, is articulated in Equation 3, with $t$ signifying the current time.

$$F_{multiStimuli}(t) = \sum_{for\ all\ stimuli\ occured\ in\ t_j} F_{singleStimuli}(t - t_j) \qquad (3)$$

Elevating the frequency of stimulation does not invariably lead to an increase in tension force. The additional influx of calcium ions becomes ineffective when all actin and myosin filaments within a MU are fully interacting. This state, termed tetanus, is formulated in Equation 4, where $MU_{MaxForce}$ denotes the maximum tension force achievable by an MU of its specific type. Thus, the dynamics of Motor Unit agents shown in Figure 4, can be characterized and modeled using Equation 4.

$$F_{MU}(t) = Minimum(F_{multiStimuli}(t)\ , MU_{MaxForce}) \qquad (4)$$

The length of a muscle is a critical factor that significantly impacts its tension force as it expresses the proportion of interacting actins and myosins in MUs. Equation 5 presents the relationship between muscle length and active tension force in which $l$ denote the length of the muscle in relation to its relaxed state known as its resting length.

$$R_{length_force}(l) = \begin{cases} 0, & l < 0.6 \\ 4l - 2.4, & 0.6 \le l \le 0.8 \\ l, & 0.8 \le l \le 1.0 \\ 1.0\,, & 1.0 \le l \le 1.2 \\ 1.0\,, & 1.2 \le l \le 1.7 \\ 3.4 - 2l, & 1.7 < l \end{cases} \qquad (5)$$

Additionally, the recruitment of a greater number of MUs by the neural system intensifies muscular tension. Consequently, the behavior of a muscle's active agent illustrated in Figure 4 can be modeled by Equation 6 where $l$ and $t$ denote the length in relation to its relaxed state of the muscle and current time, respectively.

$$F_{active}(t\ , l) = R_{length_force}(l) \times \sum_{all\ MU\ agents} F_{MU}(t) \qquad (6)$$

The passive component of a muscle produces contraction force through the inherent elasticity of the muscle. The behavior of the passive agent is formulated in Equation 7, where $F_{P0}$ is a constant specific to the type of muscle. The variable $l$, representing the muscle's length, is a relative variable that expresses the length of the muscle in relation to its relaxed state.

$$F_{Passive}(l) = Maximum\{F_{P0}\, e^{(l-1)} - 1\ , 0\} \qquad (7)$$

The force produced by a muscle is the cumulative result of its active and passive components. Consequently, Equation 8 delineates the functioning of a muscle agent. In this equation, the variables $t$ and $l$ retain the definitions provided in Equations 6 and 7.

$$F_{muscle}(l, t) = F_{Active}(l, t) + F_{Passive}(l) \qquad (8)$$

The agent-driven model has been meticulously established and derived through the physiological behavior of the muscle. Our approach, which is known as deductive reasoning, is a robust mathematical method where new insights are systematically deduced from established laws and verified facts. Consequently, the simulation outcomes predicated on this model boast a high degree of interpretability and reliability. The muscle agents can be attached to the joint agents described in Section 3.2 and, as a consequence, motion can be naturally simulated.

## 3.4. Proposed Ensemble ANN Framework for Joint-based Disorder Detection

In this study, a motion analysis method is proposed in order to differentiate between healthy and unhealthy muscles in a patient. We propose a deep learning method to investigate the output of the agent-driven simulations in each joint. Thus, neuromuscular activity around each joint can be analyzed independently. In this study, three neural networks are trained to analyze the motion in three phases. First, the raw data captured by the IoT method is used to make an initial decision. Then, using the agent-driven biomechanical model of the lower body, muscular activities are analyzed in the second phase. The neural stimulations of each muscle agent are also calculated using the proposed muscle model and are used in the third phase of pathology detection.

The proposed detection scenario is illustrated in Figure 6. As can be seen, each aspect of the motion (appearance, biomechanical, and neural aspects) is investigated by an independent model. Finally, using an averaging method, the probability of healthy or unhealthy motion in each joint is calculated. The method can be repeatedly applied to different joints so that healthy and unhealthy muscles can be differentiated.

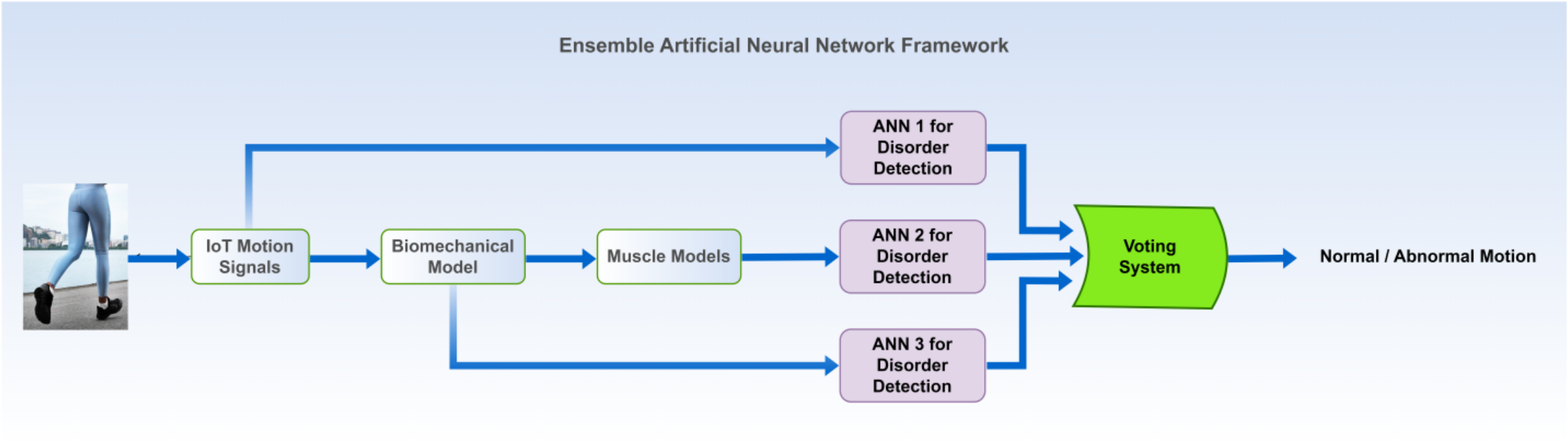

Figure 6. Proposed ensemble neural network framework for disorder detection.

The input of each neural network is a matrix including 20 floating point values containing the angles, muscular force generation, and neural stimulations during a gait cycle. As the data is complex, neural networks are used to be trained as they can understand compound patterns from the data.

In this study, a 6 layers architecture is delineated for the ANNs. The number of perceptrons in the layers are 160, 160, 160, 100, 80, and 50 respectively. Relu is considered as the activation function for the perceptrons. A person-based cross-validation method is utilized to evaluate the networks in which the test data from each person is not previously seen by the network. Our findings, which are presented in Section 4, show that the proposed method to analyze the motion is reliable.

# 4. Analysis

Our methodology includes IoT-based motion signal capturing, followed by agent-based simulation of lower body joints and muscles. These are then analyzed using an interpretable ensemble neural network framework. In this study, we aim to respond to the research questions highlighted in Section 1. In this section, we analyze our research questions to determine whether they have been satisfactorily addressed. These questions are investigated in Sections 4.1 through 4.7, respectively.

## 4.1. Reliability Analysis

In this section, the validity of the results obtained from the simulation process is examined. Since many common movement disorders, such as MS or Parkinson's, are caused by neural system disorders, calculating neuromuscular performance in these individuals is essential. To this end, a normal gait is simulated based on the proposed modeling method. In the simulation, the cycle of muscle forces generated is initially calculated based on the biomechanical model. Then, using the obtained results and the proposed agent-driven muscle model, the cycle of neural stimulations of the muscles during walking is calculated. This process implements the second and third modeling stages shown in Figure 2. Finally, the results obtained from the proposed method are compared with other researches.

In general, there are two methods for measuring neuromuscular performance. The first method, known as inverse dynamics or reverse engineering, involves calculating the patterns of muscle contractions based on kinetic characteristics such as angular changes in joints using biomechanics. A common approach is to define one equation for each joint and one variable for each muscle group [5-6]. In modeling the gait of lower limb muscles, three joints and six or seven muscle groups are assumed. As a result, the defined system of equations is unsolvable. In this case, mathematical optimization methods are used to calculate the variables and force generated by the muscles.

The second method, known as EMG, is an invasive and painful technique in which electrodes are directly inserted into the muscles to measure all electrochemical changes occurring at the electrode site. Since the electrodes in the EMG method are directly inserted into the muscles, this method is more accurate compared to the first method [74]. In this section, the results obtained from proposed movement modeling is compared with both mentioned methods as is illustrated in Figure 7, where the graphs display the neural activity of inactivity of muscle groups during a normal gait. The blue, green, and red graphs represent the results obtained from the proposed method in this paper, EMG, and inverse dynamics methods, respectively.

By comparing the results obtained from the proposed method with other methods, it can be observed that the proposed method is more accurate. For the gastrocnemius muscle, the proposed method shows 4% more overlap with EMG results compared to other inverse dynamics methods. This value is calculated to be 9%, 61%, 20%, and 37% for the tibialis anterior, hamstrings, quadriceps, and gluteus maximus muscle groups, respectively. Meanwhile, the differences between the results of the proposed method and EMG for the gastrocnemius, tibialis anterior, hamstrings, quadriceps, and gluteus maximus muscle groups are only 2%, 2%, 5%, 9%, and 20%, respectively. Since the iliopsoas muscle is a deep muscle, the electrode must be inserted deep into the muscle, which is very painful. Therefore, electromyography results for this muscle are not reported in some studies. Thus, the results for this muscle in the proposed method are only compared with other inverse dynamics methods. As a result, the comparison demonstrates the reliability of the results obtained from the proposed method.

The superiority of the proposed method over other inverse dynamics methods stems from the fact that, in this study, biomechanical equations are defined not only for each joint but also for each muscular objective. For example, one of the objectives of the hamstring muscles is to maintain the upright posture of the body at the waist. Consequently, by calculating the body's center of mass relative to the pelvis, the contraction of the hamstrings to counteract this can be calculated during specific phases of walking. As this phenomenon has not been considered in some studies, resulting in significant differences for the hamstrings and quadriceps muscles between the proposed method and other inverse dynamics methods.

The output of the proposed agent-driven simulation is the neural stimulation of the muscles during gait. In this section, the neural stimulations of the lower body muscle groups are compared with other similar inverse dynamics methods and also with EMG-based methods as the most accurate tool for neuromuscular activity detection. The periods during which neural stimuli are sent to the lower body muscle groups are depicted in Figure 7: our proposed method is illustrated with blue diagrams, EMG with green, and other inverse dynamics methods with red. For instance, EMG signals show that gastrocnemius muscle is stimulated from 8% to 54% of the gait. EMG signals are reported in [75], while inverse dynamic signals are reported in [28,76].

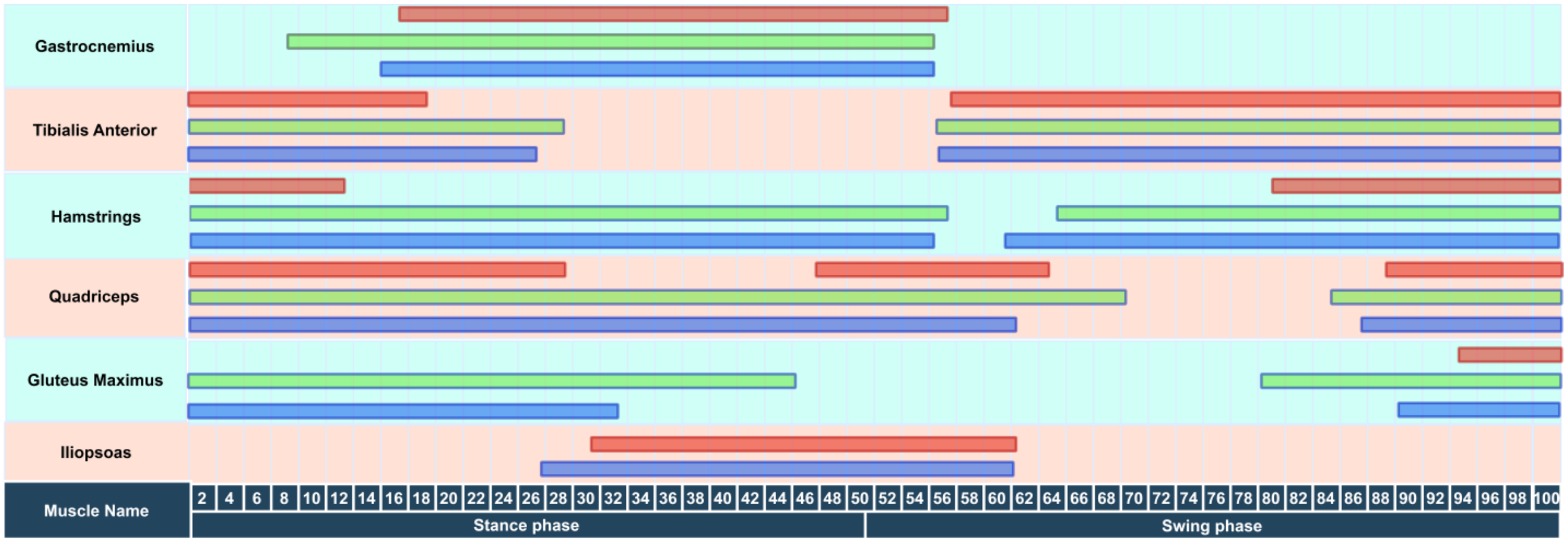

Figure 7. Comparison of neural stimulations during normal gait.

Figure 7's horizontal axis depicts a single gait cycle's stages: initiating with foot placement for the stance phase, lifting off at the cycle's midpoint to end the stance, and then foot placement again to commence the next cycle. The comparison demonstrated in Figure 7 shows that our proposed simulations for tibialis anterior, hamstrings, quadriceps, and gluteus Maximus muscles, are more reliable than other inverse dynamics methods as the results for our simulations have more overlapping with EMG signals than other inverse dynamic methods. It is considerable that the EMG methodology is the most accurate process as the electrodes are located inside the muscles, nevertheless, it is painful and invasive.

For the gastrocnemius muscle, our proposed simulation is as reliable as other inverse dynamics methods. For the iliopsoas muscle, as the muscle is in-depth, there are no EMG signals reported [75]. However, the results of our proposed simulation are as robust as other inverse dynamics methods.

## 4.2. Naturalness

As mentioned in Section 2, the naturalness of the algorithms inside the simulations is a key purpose of ABMS. This feature enhances the interpretability, accuracy, and reliability of the simulation. In this section, it is investigated how naturalness is considered in our proposed methodology.

Despite other researches that assumed EMG signals as the neural stimulation of the muscle, in this study we simulated APs as the real neural stimulations. This is the first time that an AP-based simulation is conducted in the motion analysis. EMG is an accurate estimation of electrochemical interactions inside muscles that are happened by APs, However is inherently different from stimuli. Also, MUs that are the main elements of the motion are simulated.

The proposed agent-base model of human voluntary muscle is based on the physiological structure of the muscle fibers and MUs. Consequently, the effect of environmental conditions such as temperature or fatigue cam be modeled by the proposed method. In addition, as all the voluntary muscles have same physiological and biological structure, the proposed agent-base model of the muscle can be used to simulate a wide range of muscles. Thus, we believe that the physiological and biological structure of the muscles are reflected in our proposed agent-based model and the model is also flexible.

The same is true for our proposed biomechanical model of the lower body as each joint is modeled based on mechanical principles. The proposed agent-driven biomechanical model stands for flexibility as physical changes of patients can be modeled. Our comprehensive agent-driven model of the lower body contains joint agents as well as muscle agents which is equal to how muscular and skeletal components are forming our bodies. Thereby, our agent-driven model is natural and the algorithm inside our simulation behaves exactly the same as our neuromusculoskeletal system.

## 4.3. Healthy and Unhealthy Muscles Differentiation

In this study, a patient abnormal gait analysis using the proposed method has been conducted under medical supervision. Our simulations showed that the tibialis anterior muscle in the patient was unhealthy due to the considerable changes compared to normal gait results as shown in Figure 8. However, other muscles were healthy as there were no significant changes in their results. This experiment shows that our proposed method for healthy and unhealthy muscles differentiation is reliable as the results were confirmed by clinical experts. In Figure 8, neural stimulation of the muscle during a gait cycle is simulated in which the vertical axis represents the number of APs, and the horizontal axis shows the elapsed time of the gait cycle beginning from the stance phase.

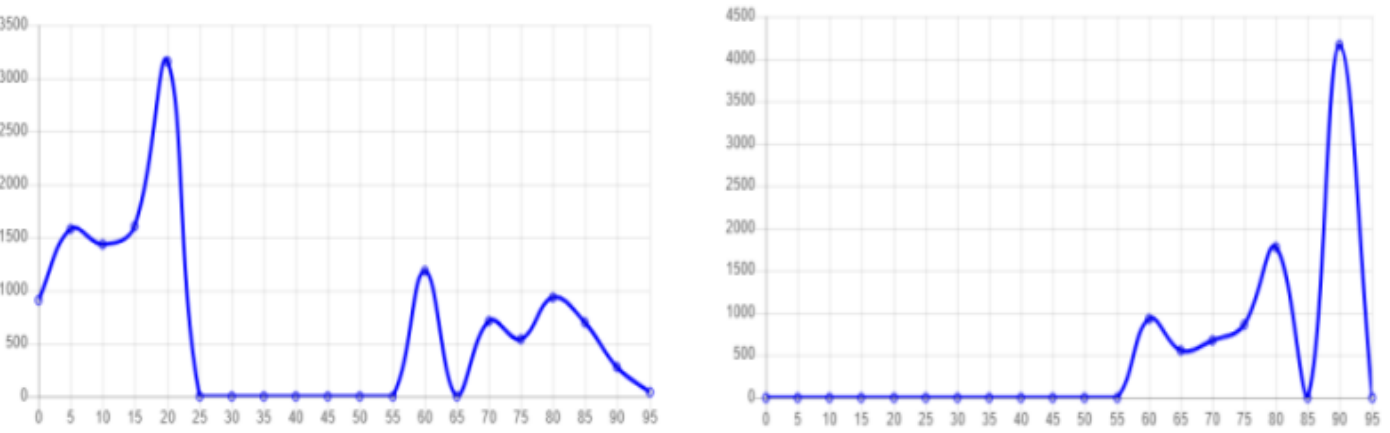

(a) (b)

Figure 8. Comparison between results of gait simulation for healthy (a) and unhealthy (b) tibialis anterior muscle.

The changes observed in the tibialis anterior muscle were not limited to neural stimulations; these changes also included raw movement signals obtained through the IoT-based method and the output of the biomechanical model, which represents muscle force patterns. All these observations indicated a disorder in the tibialis anterior muscle. Since no significant changes were observed in other joints, it was concluded that the other muscles in that patient were healthy. As this research was conducted under the supervision of clinical specialists, the results were also confirmed by clinical specialists, indicating the reliability of the proposed method for distinguishing between healthy and unhealthy muscles. Based on this research, we suggested that to differentiate between healthy and unhealthy muscles, the examination and analysis of movement should be conducted separately for each joint. Any observed disorder in a joint should lead to a more detailed examination and study of the muscles associated with that joint.

Generally, we proposed a joint-based strategy in which muscle groups can be analyzed independently. As a consequence, unhealthy muscles can be distinguished from healthy muscles in the treatment protocols and medications. Thus, the harmful effects of medications on healthy muscles can be minimized.

## 4.4. User-friendly Interface Analysis

A user-friendly graphical interface is an important factor in software development as a powerful application with many computations is not valuable if it cannot be used easily by users. In this study, modern graphical interfaces are used to enhance the interface of the proposed application.

A User-friendly graphical interface facilitates the learning of the software. For instance, while OpenSim is a well-known tool for musculoskeletal modeling and simulation, some users have reported challenges with its graphical user interface. These challenges include the steep learning curve for new users and the need for improvements in user-friendliness and visualization capabilities [77]. Despite the OpenSim which contains a high number of windows, buttons, menus, and other interface components making it complex to use, our proposed application includes a surprisingly smaller number of three buttons in a single simple window which promotes the usability of the software.

Some screenshots of the proposed application are demonstrated in Figure 9. As can be seen, the user can capture his motion signals by clicking a button on the application (Figure 9.a) and then, be informed whether his motion in the joint of interest is normal or abnormal (Figures 9.b and 9.c), and also simulate his own captured motion on his mobile by a 3D animation (Figure 9.d). The application can also be executed on desktop computers as illustrated in Figure 9.e. For inexpert users, the application is easy to use as the user can start working by only clicking a button and the results from the proposed neural network framework are displayed in an easy-to-understand graphical interface. The interface is powerful as even an inexpert user can learn and use it easily.

The simplicity of the graphical user interface of the proposed tool stems from minimizing the number of interactive graphical elements, such as buttons. This feature allows users to quickly learn how to use the tool. The 3D animation display of the user's simulated movement, generated using their movement signals, along with simple and self-explanatory images, and the absence of overwhelming graphical components like texts and windows, facilitate the software's learning process. This reduces or eliminates the need to refer to software documentation or manuals.

The importance of this feature can be examined from the perspectives of both specialist and non-specialist users. For specialists working in clinical settings, it is essential to consider the stress and psychological pressure they face, which can unintentionally reduce their accuracy. Therefore, the user interface should be as simple as possible to fill this gap. For non-specialist users, a simple user interface increases the tool's accessibility to the public. Consequently, non-specialists can use the proposed method to analyze their movements at minimal cost. This approach enhances individual health awareness in society, enabling those with disorders to identify and manage them more quickly.

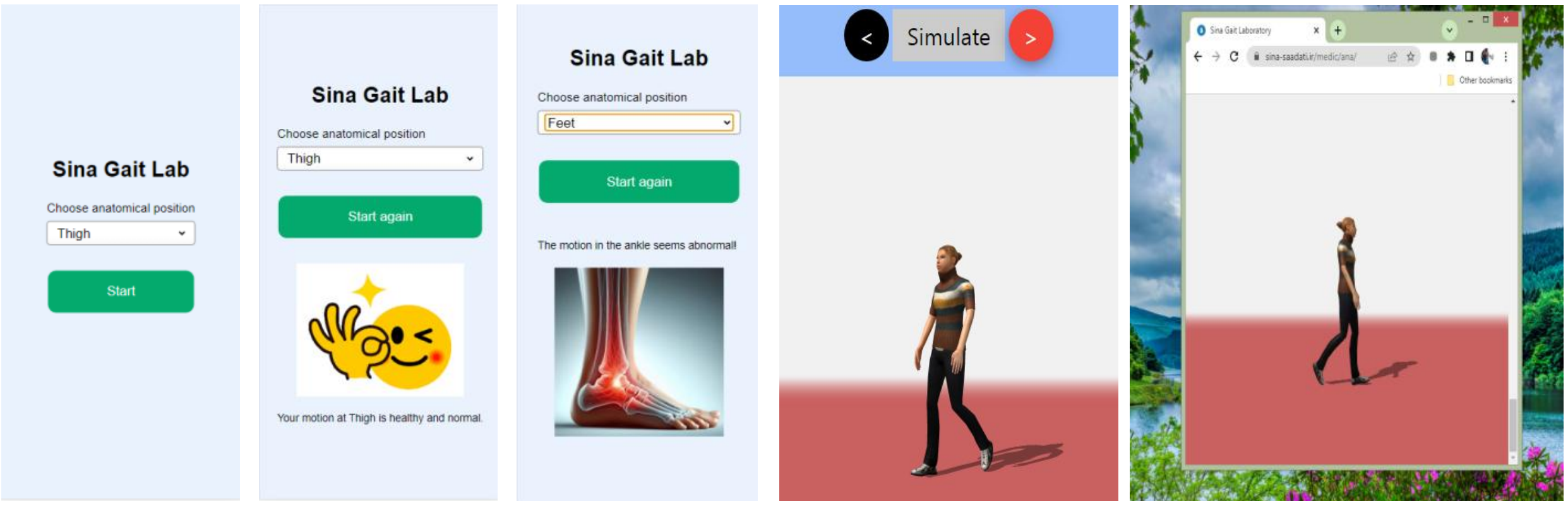

(a) ( b) (c) (d) (e)

Figure 9. Graphical Interface of the proposed web-based application.

## 4.5. Cloud Computing Analysis

Cloud computing is utilized in this study to develop an infrastructure-independent application which can be executed on a wide range of computers. However, the facilities of a cloud-based application are not restricted in infrastructure as cloud computing provides software with a higher level of availability and minimizes energy consumption at the same time, resulting in green computing.

In this study, running the proposed web-based application was evaluated on computers with different hardware and software infrastructures. Contrasting with the approach outlined in [63], where a cloud-based platform application is developed with local processing, our proposed application is entirely cloud-based. This design ensures that all computational tasks are offloaded to the cloud, fostering an infrastructure-independent software that is globally accessible. Moreover, this approach aligns with the principles of green computing, as it leverages the cloud's potential for energy efficiency and reduced environmental impact.

## 4.6. IoT Computing Analysis

In this paper, as illustrated in Figure 2, a gait modeling method for investigating and analyzing neuromuscular performance is proposed. The second and third stages of the proposed method were examined for reliability in Section 4.1, where the simulation input, including the angular changes of the lower limb joints during a gait cycle, was gathered from other studies. This is because Section 4.1 focused solely on the presented model of the neuromuscular skeletal system to assess the reliability of the simulation results. In this section, we will examine the first stage in Figure 2, which involves evaluation of obtaining movement signals using the three-axis compass sensor of smartphones.

To digitalize human gait using smartphones, it is important to note that the body's movement signals during walking, which include angular changes in the lower limb joints, will vary for different steps and individuals. However, what remains constant in natural walking are not the exact signal values but the patterns that can be used to assess the health of each joint [28]. Therefore, in this study, participants were asked to take several steps instead of just one. The signal obtained for each step taken is then broken down into several graphs. Finally, using regression techniques, a single graph is calculated as a summary of all the signals received during different steps taken. This method is repeated for each joint, resulting in a graph for each joint that represents the joint angles in a gait cycle. The process is demonstrated in Figure 3.

The joint angle charts during a gait cycle were compared with previous charts from other studies. The differences between the charts were negligible, indicating that the proposed method for storing motion signals is sufficiently accurate and reliable. However, using musculoskeletal neuromuscular modeling, muscle contraction force charts and neural stimulation charts for different muscle groups were calculated and compared with previous results. Again, no significant differences were observed. It is worth mentioning that the charts obtained in similar research were collected using wearable sensors, whereas in this study, smartphones were used for this purpose. The importance of using smartphones instead of wearable sensors lies in their ubiquity. Smartphones are accessible in much more parts of the world, even in many deprived countries and villages, whereas wearable sensors are hard to obtain even in some major cities like Tehran. Consequently, using smartphones can significantly facilitate the motion modeling process in terms of cost and time.

In this study, we introduce an innovative IoT-based method for capturing motion signals, marking its first application in this field. For evaluating the approach, the raw data obtained from our IoT-based motion-capturing application were compared with traditional markers and insole-based signal-capturing tools, under the supervision of clinical experts. Our findings indicate that the sensors embedded in smartphones are as precise as conventional tools, as evidenced by the congruence of the raw data from our application with other signals. However, our proposed method offers enhanced accessibility and cost-efficiency. From availability and infrastructure independence points of view, our IoT-based application was tested on mobiles with different hardware and software platforms. In this case, the application was successfully executed on Sony, Samsung, Xiaomi, and iPhone mobiles with iOS and Android operating systems.

In addition, we used our IoT application to classify each phase of the gait cycle for our captured instances as detailed in Table 3. Compared to existing research, our proposed methodology stands out for its enhanced interpretability and consequent accuracy improvements. For instance, while studies [68-69] have relied on data-driven machine learning models that entail substantial computational overhead and complexity, our study leverages a novel IoT application to precisely capture the angles of lower body joints. This direct measurement allows for straightforward classification of gait phases by calculating the feet's position. Our empirical results demonstrate a significant reduction in computational load while achieving superior accuracy, making our approach not only more interpretable but also more efficient and practical for real-world applications.

Table 3. Comparison of accuracy for different gait phase classifiers.

| Method | [68] | [69] | Proposed IoT Method |
|---|---|---|---|
| **Accuracy** | 99.9 | 90.6 | 100.0 |

## 4.7. Deep Learning Evaluation

In this section, the fourth stage of the proposed method, which includes a deep learning-based ensemble framework for detecting healthy or unhealthy gait as shown in Figure 2, is evaluated. In this study, 6 ANN models are developed that analyze different aspects of motion during gait independently and detect intelligently whether the motion is normal. A person-based cross-validation evaluation strategy is followed in which the data originating from a person is only seen by the networks once. The experiments are detailed in Table 4 in which for each joint, motion signals from the proposed IoT-based method, muscular contraction forces, and neuromuscular stimulation are processed in experiment1, 2, and 3, respectively.

Table 4. Accuracy of ANN models for joint-based normal/abnormal motion detection.

| Joint of interest | Experiment 1 | Experiment 2 | Experiment 3 | Average | |
|---|---|---|---|---|---|
| Hip Joint | 97.78 | 98.32 | 93.17 | 96.42 | 96.48 |
| Knee Joint | 95.93 | 99.64 | 94.02 | 96.53 | |

In this study, different ANN models are designed to analyze one joint motion. As each model concentrates on only one aspect of the motion, we believe that the interpretability of the proposed method is enhanced compared to other researches. For designing the architecture of the ANN models, a trial and test scenario is pursued. The same scenario is followed in [78] in which one feature is analyzed in each step and the case with the highest accuracy is selected to be checked in further tests.

In this scenario, different numbers of hidden layers are investigated in the first step from one to ten hidden layers. Then, for the best case, different numbers of perceptrons are investigated for each layer in the second step. This step was time-consuming as the number of needed tests for a network with 6 hidden layers is 6×20 and the discrepancy between the number of perceptrons for each continuous step is considered as 10. Finally, 4 mathematical functions are investigated as the activation function of the networks including ReLu, Tanh, identity, and logistic. Figure 10 details our methodology for architectural design.

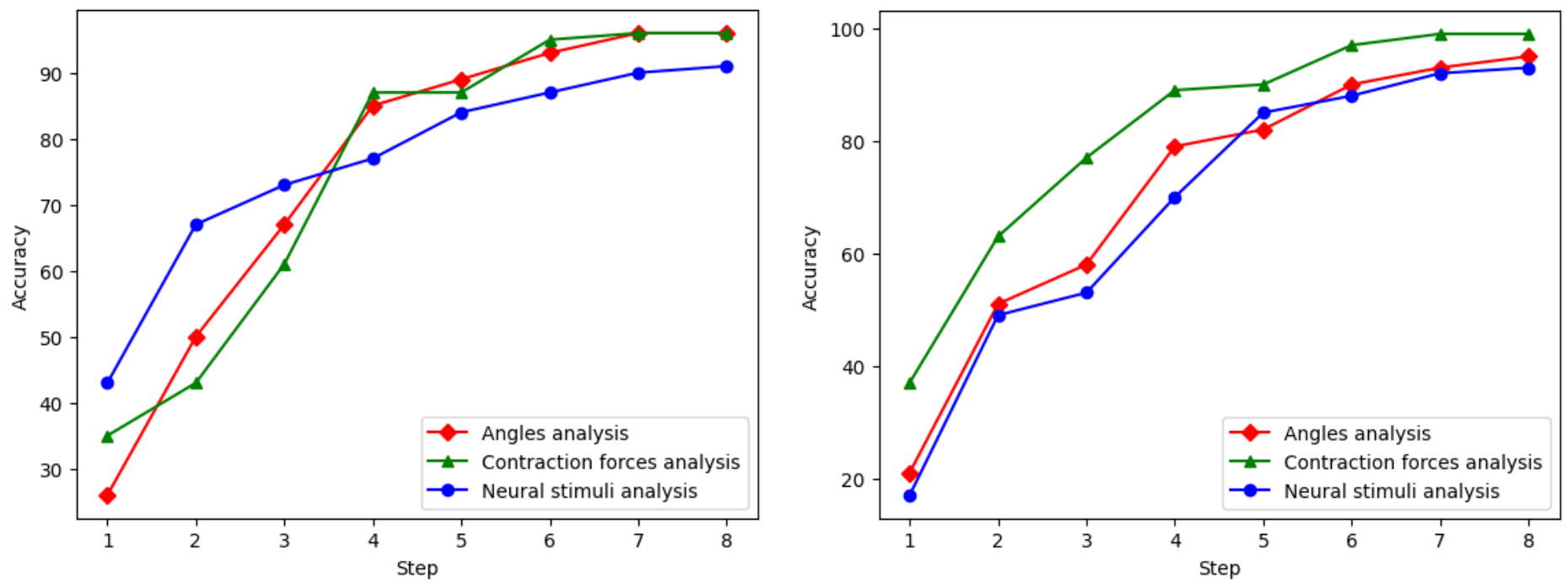

(a) (b)

Figure 10. Accuracy promotion of ANN models for the hip joint (a) and knee joint (b).

Despite other deep learning methods that contain a big ambiguous neural network in which different aspects of the motion are analyzed, in this study for each joint in the lower body, an ensemble neural network framework is proposed in which different independent networks investigate different aspects of the motion including the appearance, muscular forces, and neural stimulations. This strategy is also equivalent to what a physician does to detect a motion disorder. Therefore, we believe that the proposed deep learning method supports the interpretability.

Using an ensemble neural network framework not only promotes the interpretability of the system but also enhances the accuracy and reliability of the system. As mentioned in Section 2.3, in this architecture, different networks tend to learn the same patterns while learning different noises. Thus, in a system containing multiple networks for each detection, noises are ignored.

## 5. Conclusion

In this study, a method is presented to analyze human motion to differentiate between healthy and unhealthy muscles in a patient. To facilitate the motion signal capturing process, an IoT-based methodology is introduced in which smart mobiles can be used to digitalize the human motion. Then, an agent-driven biomechanical model of the human lower body is introduced to simulate the gait cycle for an inverse dynamic calculation in which angular displacements of the lower body joints of a patient captured by the smart mobiles are the input of the simulation and the contraction forces of the lower body muscle groups are the outputs. In the third step, our proposed agent-based model of the muscle is used to calculate the neural stimuli entered to each muscle group from the contraction force. The output is the number of APs entered to MUs of the muscles.

Our proposed agent-driven simulation contain biomechanical agents of the lower body joints and muscular agents which are connected to the joint agents. From a high level, this simulation reflects the neuromusculoskeletal structure of the body and from a low level, the biomechanical, physiological, and biological aspects of the motion are meticulously considered in our simulation. Therefore, our agent-driven simulation stands at a high level of scientific interpretability and, as a consequence, reliability. The results of the simulation can be directly analyzed by medical experts, however, an interpretable ensemble deep learning framework is developed in this study to assist clinicians to joint-based motion abnormally detection.

Our proposed analyzing method roots in the scientific facts in order to promote the interpretability and reliability of the process. However, software techniques are used to enhance the usability of the proposed motion analysis method. First, our proposed IoT-based method for capturing the motion signals by smart mobiles, which is the first of its kind, promotes the accessibility and usability of the process as smart mobiles are worldwide available, even for people in poor regions. Second, cloud computing is utilized to develop an infrastructure independent application that can be executed on a wide range of computers, from desktops to mobiles and tablets. In addition to the platform independency, cloud-based platform makes the application globally available. Third, our proposed user-friendly graphical interface facilitates working with the software as it is easy-to-learn and easy-to-use, even for inexpert users.

As conclusion, our proposed methodology paid attention to both scientific aspects of simulation and software requirements. Thus, the proposed application stands at a high level of interpretability, reliability, and usability. As a consequence, we hope that the proposed methodology pave the way of clinical processes for both clinicians and patients.